\documentclass[aps,prl,reprint,twocolumn,longbibliography,nobalancelastpage,superscriptaddress,notitlepage]{revtex4-1}
\usepackage{amsmath,amssymb,amsfonts}
\usepackage[pdftex]{graphicx}
\usepackage{color}
\usepackage{braket}
\usepackage[colorlinks=True,linkcolor=blue,citecolor=blue,urlcolor=blue]{hyperref}
\usepackage{hypcap}

\newcommand{\pd}{{\phantom\dagger}}
\begin{document}

\title{Exact solution of a percolation analogue for the many-body localisation transition}

\author{Sthitadhi Roy}
\email{sthitadhi.roy@chem.ox.ac.uk}
\affiliation{Physical and Theoretical Chemistry, Oxford University, South Parks Road, Oxford OX1 3QZ, United Kingdom}
\affiliation{Rudolf Peierls Centre for Theoretical Physics, Clarendon Laboratory, Oxford University, Parks Road, Oxford OX1 3PU, United Kingdom}

\author{David E.~Logan}
\email{david.logan@chem.ox.ac.uk}
\affiliation{Physical and Theoretical Chemistry, Oxford University, South Parks Road, Oxford OX1 3QZ, United Kingdom}

\author{J.~T.~Chalker}
\email{john.chalker@physics.ox.ac.uk}
\affiliation{Rudolf Peierls Centre for Theoretical Physics, Clarendon Laboratory, Oxford University, Parks Road, Oxford OX1 3PU, United Kingdom}

\begin{abstract}
We construct and solve a classical percolation model with a phase transition that we argue acts as a proxy for the quantum many-body localisation transition. The classical model is defined on a graph in the Fock space of a disordered, interacting quantum spin chain, using a convenient choice of basis. Edges of the graph represent matrix elements of the spin Hamiltonian between pairs of basis states that are expected to hybridise strongly. At weak disorder, all nodes are connected, forming a single cluster. Many separate clusters appear above a critical disorder strength, each typically having a size that is exponentially large in the number of spins but a vanishing fraction of the Fock-space dimension. We formulate a transfer matrix approach that yields an exact value $\nu=2$  for the localisation length exponent, and also use complete enumeration of clusters to study the transition numerically in finite-sized systems. 
\end{abstract}

\maketitle

Insights into quantum many-body systems can be gained at a variety of levels from studying classical problems. An exact equivalence is provided by the well-known mapping between ground-state properties of a quantum system and finite-temperature behaviour in  classical system in one higher dimension~\cite{suzuki1976relationship,sachdev2011quantum}. Qualitative understanding in quantum systems may however be derived from classical models in other ways, phenomenological in nature and rooted in physical argument. For example, the analogy between Anderson localisation and classical percolation provides a picture which is particularly useful in the context of the integer quantum Hall effect~\cite{tsukada1976tail,kazarinov1982quantum,trugman1983localization}. 
Many-body localisation transitions \cite{basko2006metal,gornyi2005interacting,oganesyan2007localisation,znidaric2008many,pal2010many,luitz2015many,kjall2014many,nandkishore2015many,imbrie2016many} are currently of great interest. They
occur in highly excited states and concern dynamical properties of quantum systems. They are therefore not generally expected to admit exact mappings onto classical problems. 
The question of whether or how classical statistical mechanical models can be constructed that mimic aspects of such phase transitions is thus naturally of fundamental interest.

\begin{figure}
\includegraphics[width=\columnwidth]{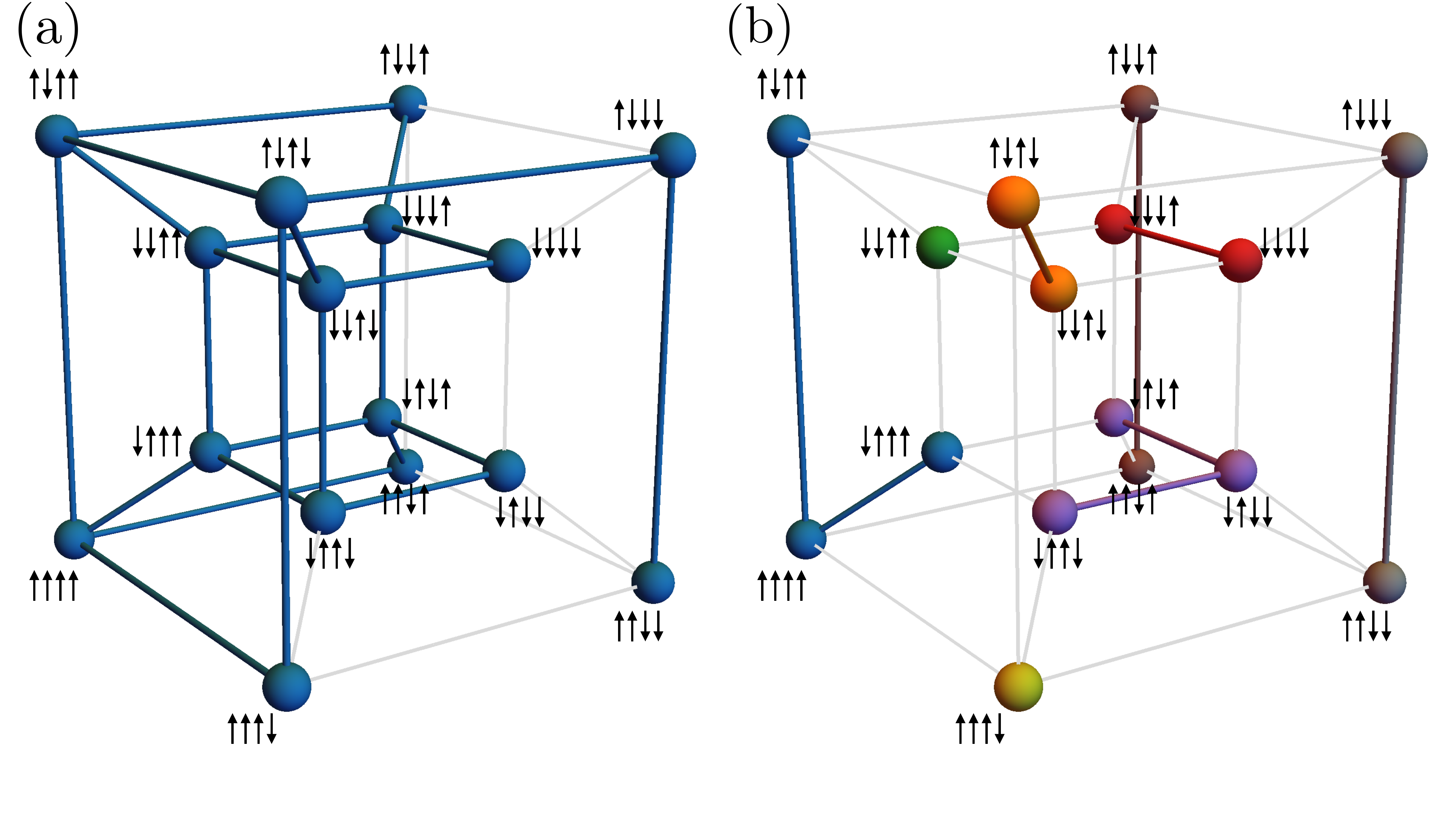}
\caption{Caricatures of (a) the percolating (delocalised) and (b) the non-percolating (localised) phases for a system of four spins-1/2. The nodes represent the $2^4=16$ basis states of the Fock space as indicated by the spin configurations. All nodes in a given cluster, and the \emph{active} edges joining them, have the same colour. Edges coloured grey are \emph{inactive}: they represent non-zero matrix elements in the quantum Hamiltonian that fail to satisfy the percolation criterion of Eq.~\eqref{eq:percolationcriterion}.}
\label{fig:hslattice}
\end{figure}

In this paper we formulate a classical percolation problem inspired by the quantum mechanics of a disordered spin system. The percolation problem is defined on a graph in the Fock space of the spin chain.  We find that there exists a percolation transition~\cite{stauffer2014introduction} in the classical model which mimics certain aspects of the many-body localisation transition in the quantum model. In particular, our choice of diagnostic for the percolation transition is motivated by the behaviour across the many-body localisation transition of the Fock-space participation entropies of eigenstates of the quantum system~\cite{deluca2013ergodicity,luitz2015many}. 

Remarkably, the classical percolation problem arising in this way from a commonly-studied spin model with local interactions, admits an exact solution using a transfer matrix. This allows us to extract analytically the critical disorder strength and the value $\nu=2$ for the localisation length exponent. In addition, we corroborate and extend our results by enumerating Fock-space clusters exactly for finite-sized systems. Our approach is complementary to works that use a phenomenological renormalisation group~\cite{vosk2015theory,potter2015universal,dumitrescu2017scaling,goremykina2018analytically,dumitrescu2018kosterlitz} or the semiclassical limit of Clifford circuits~\cite{chandran2015semiclassical} to construct a classical percolation problem in real space, in that we work entirely in Fock space (as was done in Ref.~\cite{logan2018many}) but use classical rules applied to individual realisations of the microscopic quantum model.

We first discuss the construction of the classical problem. The Hamiltonian of a quantum many-body system can be expressed as a tight-binding model in Fock space~\cite{welsh2018simple}, with the form
\begin{equation}
\mathcal{H} = \sum_{I}\mathcal{E}_I^\pd\ket{I}\bra{I}+\sum_{I\neq K}T_{IK}^\pd\ket{I}\bra{K},
\label{eq:tightbinding}
\end{equation}
where $\{\ket{I}\}$ denotes a set of many-body basis states. We consider a graph in Fock space, consisting of nodes that represent these basis states and edges that indicate non-zero matrix elements $T_{IK}$. A suitable choice of basis is the one corresponding asymptotically to eigenstates in the strong disorder limit.

The percolation problem we study arises by designating edges \emph{active} or \emph{inactive} according to a microscopically based rule. 
An edge is defined to be active iff
\begin{equation}
\vert T_{IK}\vert > \vert \mathcal{E}_{I}-\mathcal{E}_{K}\vert.
\label{eq:percolationcriterion}
\end{equation}
This rule is motivated by the fact that for a two-state quantum mechanical problem consisting of energy levels with separation $\Delta$ and coupling $J$, the extent of hybridisation is controlled by the ratio $J/\Delta$. A percolation cluster is a maximal set of nodes joined by edges that are active, and a transition occurs in the model we treat because a decreasing proportion of edges are active as disorder strength is increased.

An appropriate diagnostic for the phase transition in the classical model can be defined as follows, taking motivation from the behaviour of participation entropies of quantum mechanical eigenstates. For a many-body eigenstate $\ket{\psi}$ the first participation entropy in Fock space is $S_1=-\sum_{I}\vert \braket{\psi|I}\vert^2\log \braket{\psi|I}\vert^2$. Denoting the Fock-space dimension by $N_{\cal H}$, a characteristic feature of the delocalised phase is the scaling $S_1\sim a_1 \log N_\mathcal{H}$ with $a_1=1$. By contrast, in the localised phase, $a_1 < 1$~\cite{luitz2015many}.
In a classical model, since there are no probabilities of the form $\vert\braket{\psi|I}\vert^2$, we define a scaled indicator function
\begin{equation}
p_I = \begin{cases}
		1/N_\mathcal{C} ~~~: I\in \mathcal{C}\\
		0 ~~~~~~~~:\mathrm{otherwise}\,,
	  \end{cases}
\label{eq:indicator}
\end{equation}
where $N_\mathcal{C}$ denotes the number of nodes in the cluster $\mathcal{C}$.
The distribution of $p_I$ over Fock space plays an analogous role to probabilities derived from a quantum wavefunction, 
and the equivalent of the participation entropy is
\begin{equation}
\overline{-\sum_{I}p_I\log p_I} = \overline{\log N_\mathcal{C}}=\log \mathcal{S}_{\mathrm{typ}}.
\label{eq:Styp}
\end{equation}
This is simply the logarithm of the typical cluster size $\mathcal{S}_\mathrm{typ}$, with an average over disorder realisations denoted by $\overline{(\cdot)}$. 
For completeness we also define the average cluster size $\mathcal{S}_\mathrm{avg} = \overline{N_\mathcal{C}}$.
In analogy with participation entropies, we expect $\mathcal{S}_\mathrm{avg/typ} \sim N_\mathcal{H}^{\alpha_\mathrm{avg/typ}}$ with $\alpha_\mathrm{avg/typ} = 1$ in the percolating (delocalised) phase and $\alpha_\mathrm{avg/typ} < 1$ in the non-percolating (localised) phase. Hence we use the value of $\alpha_\mathrm{avg/typ}$ as a diagnostic for the phase transition.

\begin{figure}
\includegraphics[width=\columnwidth]{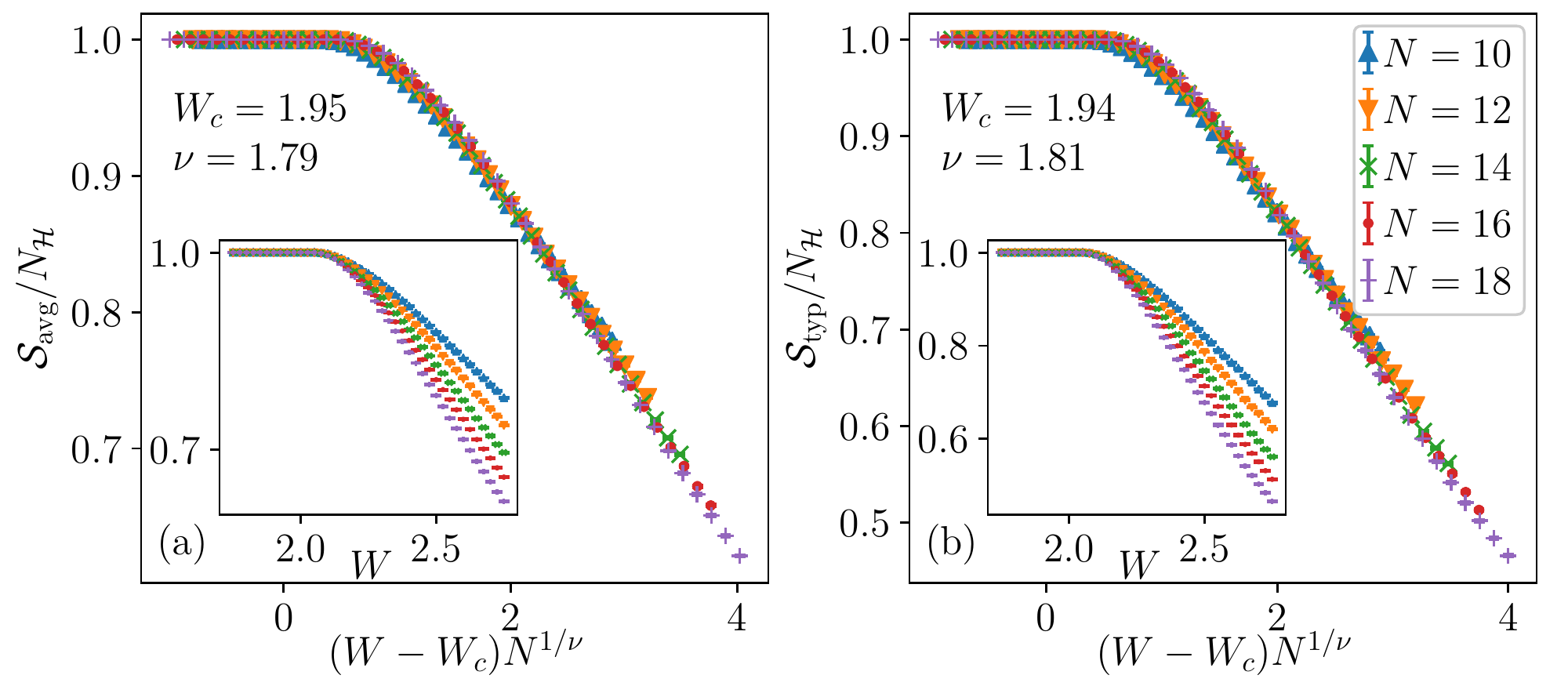}
\caption{Average (left panel) and typical (right panel) cluster sizes, normalised by Fock-space dimension, as a function of disorder strength for various system sizes $N$. Insets show raw data and main panels show scale-collapsed data with best fits for $W_c$ and $\nu$. Calculations use $J_z=1$ and $J=4.1$ with $10^5$ realisations and errors estimated via the standard bootstrap method with 500 resamplings.}
\label{fig:numerics}
\end{figure}

To put the entire formulation on a concrete footing we consider a quantum Ising chain of spins-1/2 with disordered longitudinal and uniform transverse fields, described by the Hamiltonian
\begin{equation}
\mathcal{H} = J_z^\pd\sum_{\ell=1}^{N-1}\sigma^z_\ell\sigma^z_{\ell+1} + \sum_{\ell=1}^Nh_\ell^{\pd}\sigma^z_\ell + J\sum_{\ell=1}^N\sigma^x_\ell,
\label{eq:hamising}
\end{equation}
where $\sigma_{l}^{z} =\pm 1$, and $h_\ell\in[-W,W]$ are random fields drawn from a uniform distribution.
As the disorder couples to the $\{\sigma^z_\ell \}$, a natural choice for $\{\ket{I}\}$ is the basis composed of product states with $\sigma^z_\ell=\pm1$. Since the off-diagonal part of $\mathcal{H}$ is simply $\sum_\ell\sigma^x_\ell$, the Fock-space graph is precisely an $N$-dimensional hypercube with edges between pairs of spin configurations that differ only on one spin, as illustrated in Fig.~\ref{fig:hslattice}. The Fock-space node energies $\mathcal{E}_I=\bra{I}J_z^\pd\sum_{\ell}\sigma^z_\ell\sigma^z_{\ell+1}+h_\ell^\pd\sigma^z_\ell\ket{I}$ can be evaluated straightforwardly with our basis choice, and the energy difference $\vert \mathcal{E}_{I}-\mathcal{E}_{K}\vert$ between states connected by spin reversal at site $\ell$ is  $2\vert J_z^\pd(\sigma^z_{\ell+1}+\sigma^z_{\ell-1})+h_\ell^\pd\vert$. This includes both an on-site term and cooperative contributions that depend on the states of neighbouring spins. To ensure that all edges are active in the weak disorder limit ($W\to 0$) we require $J>4J_z$. 

Before presenting our analytic treatment of this model, we show numerical results for $\mathcal{S}_\mathrm{avg/typ}$ obtained by exact enumeration of clusters in a finite system \footnote{For definiteness, we consider the cluster that contains the node $I_0$ with $\mathcal{E}$ closest to the mean, so that $\mathcal{E}_{I_0} = \min_{I}\{\vert \mathcal{E}_I-\bar{\mathcal{E}}\vert\}$ where $\bar{\mathcal{E}} = \sum_{I}\mathcal{E}_I/N_\mathcal{H}$.}. From Fig.~\ref{fig:numerics}, it is clear that there is a critical disorder strength $W_c$. For $W<W_c$ we find $\mathcal{S}_\mathrm{avg/typ}\sim N_\mathcal{H}$, indicating a percolating phase. For $W>W_c$ we find a localised phase in which $\mathcal{S}_\mathrm{avg/typ}\sim N^{\alpha_{\mathrm{avg/typ}}}_\mathcal{H}$ with $\alpha<1$. 
The latter scaling is akin to the behaviour of quantum eigenstates in the many-body localised phase~\cite{deluca2013ergodicity,luitz2015many}.
The values of $W_c$ and $\nu$ can be obtained by collapsing the data for various system sizes onto a universal function $g[(W-W_c)N^{1/\nu}]$. The data for $\mathcal{S}_\mathrm{avg}$ yields $W_c\approx 1.95$ and $\nu\approx 1.79$ whereas those for $\mathcal{S}_\mathrm{typ}$ give $W_c\approx 1.94$ and $\nu\approx 1.81$. These results are close to the exact values $W_c= J/2=2.05$ and $\nu=2$ derived below.

An important further numerical observation is that in the percolating phase \emph{all} Fock-space nodes form a single cluster, as indicated by the fact that $\mathcal{S}_\mathrm{avg/typ}$ not only scales linearly with $N_\mathcal{H}$ but is equal to it. This behaviour is completely different from that for bond percolation on a finite-dimensional lattice. It arises because of the high coordination number ($=N$) of the Fock-space graph, and despite the fact that for $W\lesssim W_c$ a finite fraction of edges are inactive. Both the formation of a single cluster and the presence of inactive edges in the percolating phase are central to our analytical approach, which we sketch below with further details in \cite{supp}.

To indicate the existence of a single cluster, we consider in a system of size $N$ a function of the disorder realisation with the properties
\begin{equation}
X_N(\{h_\ell\}) = \begin{cases}
					1 ~~~: \mathrm{if}~ N_\mathcal{C}=N_\mathcal{H},\\
					0 ~~~: \mathrm{otherwise}.
				  \end{cases}
\label{eq:XNdef}
\end{equation}
Our strategy is to relate systems with open boundary conditions of size $N$ and $N+1$, by writing an expression for $X_{N+1}$ in terms of $X_N$ and $h_{N+1}$. This constitutes our basic recursion relation. We then perform a disorder average over all the fields and solve the averaged recursion relation to obtain the probability that all the nodes of the Fock-space graph belong to the same cluster.
In this way we identify the critical point and determine the localisation length exponent.

\begin{figure}
\includegraphics[width=0.8\columnwidth]{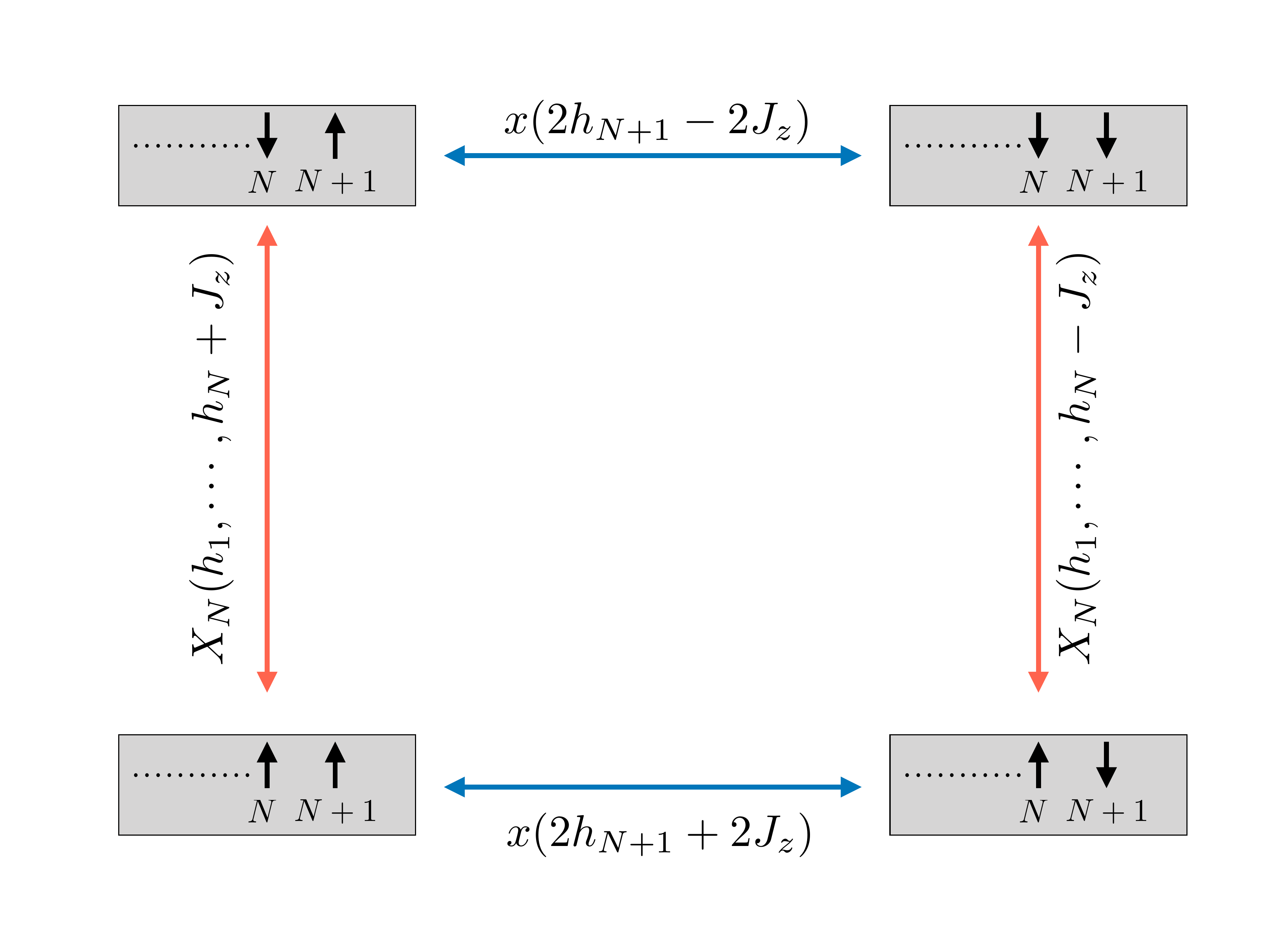}
\caption{Schematic representation of transitions involving spin flips at sites $N$ and $N+1$.}
\label{fig:4Dstatespace}
\end{figure}

To find a recursion relation, we must account for two points. First, the edge representing a spin flip at site $N+1$ may be active or inactive, depending both on the value of $h_{N+1}$ and (via the exchange interaction) on the spin orientation at site $N$. Second, the effective field $h_N$ acting on the spin at site $N$ is modified in the $N+1$ site system by the exchange interaction with the spin at site $N+1$. To represent these features we define the indicator function 
\begin{equation}
x(\Delta) = \begin{cases}1~~~:\vert \Delta\vert<J,\\0~~~:\mathrm{otherwise},\end{cases}
\label{eq:xn}
\end{equation}
where $\Delta={\cal E}_I - {\cal E}_K$ for configurations $|I\rangle$ and $|K\rangle$ connected by a given spin flip.
Transitions on the Fock-space graph arising from spin flips at sites $N$ and $N+1$ are shown schematically in Fig.~\ref{fig:4Dstatespace}. 
Those denoted with blue arrows arise from spin flips at site $N+1$ and are influenced by the exchange field only from the spin at site $N$. Hence $\Delta = 2h_{N+1}+ 2J_z \sigma^z_N$ in this case. Transitions denoted with red arrows arise from spin flips at site $N$ and have $\Delta = 2(h^\pd_{N}+J^\pd_z\sigma^z_{N+1})+ 2J_z^\pd \sigma^z_{N-1}$. Here the exchange interaction with the spin at site $N+1$ is equivalent to the modification $h^\pd_N\to h^\pd_{N}+J^\pd_z\sigma^z_{N+1}$ of the field on site $N$. 

In order that the four nodes represented in Fig.~\ref{fig:4Dstatespace} are connected, it is sufficient that at least three of the edges depicted are active. From this we derive the recursion relation
\begin{align}
X^\pd_{N+1} = &X_{N}^+X_{N}^-(x_{N+1}^+ + x_{N+1}^- -3x_{N+1}^+x_{N+1}^-)+\nonumber\\
&(X_{N}^+ + X_{N}^-)x_{N+1}^+x_{N+1}^-.
\label{eq:XNrec}
\end{align}
Here $X_N^\pm = X_N(h_1,\cdots,h_N\pm J_z)$ and $x_N^\pm = x(2h_N\pm 2J_z)$.
To obtain a closed set of equations we require expressions for $X^{\pm}_{N+1}$. These are simple to obtain because in Eq.~\eqref{eq:XNrec} the field $h_{N+1}$ enters the expression for $X_{N+1}$ only via $x_{N+1}^\pm$, which is independent of the fields at other sites. Hence equations for $X_{N+1}^\pm$ are obtained by replacing $h_{N+1}$ with $h_{N+1}\pm J_z$ in the argument of $x_{N+1}^\pm$.
Defining the notation $z^\pm_N = x(2h^\pd_N\pm4J_z)$, $z_N=x(2h_N)$ and $Y^\pd_N(\{h_\ell\})=X_N^+X_N^-$, the recursion relations for $X_{N+1}^\pm$ are
\begin{align}
X_{N+1}^\pm =& Y_N^\pd(z_{N+1}^\pd+z_{N+1}^\pm-3z_{N+1}^\pd z_{N+1}^\pm)+\nonumber\\&(X_N^+ + X_N^-)z_{N+1}^\pd z_{N+1}^\pm.
\label{eq:XNpmrec}
\end{align}
Finally, we can find an expression for $Y_{N+1}$ from Eq.~\eqref{eq:XNpmrec} by multiplying the recursions relations for $X_{N+1}^+$ and $X_{N+1}^-$, giving
\begin{align}
Y_{N+1}^\pd = &Y_N^\pd[z_{N+1}^\pd+z_{N+1}^+z_{N+1}^- - 3z_{N+1}^\pd z_{N+1}^+z_{N+1}^-] + \nonumber\\
&(X_N^+ + X_N^-)z_{N+1}^\pd z_{N+1}^+z_{N+1}^-.
\label{eq:YNrec}
\end{align}
Linearity in $Y_{N+1}$ and $X_{N}^\pm$ is retained under multiplication because $Y_N$ and $X_N^{\pm}$ are idempotent.

The recursion relations \eqref{eq:XNpmrec} and \eqref{eq:YNrec} are a closed set of coupled linear equations for $Y_N$ and $X_N^\pm$. They have the form $\mathbf{v}_{N+1}=\mathbf{M}_{N+1}\mathbf{v}_N$ where $\mathbf{v}^\pd_N=(Y_N^\pd,X_N^+,X_N^-)^T$.
Crucially, the matrix $\mathbf{M}_{N+1}$ depends only on $h_{N+1}$ and not on $h_\ell$ for $\ell \leq N$.
Due to this and the linearity, the system of equations can be converted into one for the averaged quantities. Defining the disorder average by $\overline{(\cdot)}=\left(\prod_{\ell=1}\int dh_\ell P(h_\ell)\right)(\cdot)$ where $P(h)=\Theta(W-\vert h\vert)/2W$, the averaged system of equations is
\begin{equation}
\overline{\mathbf{v}}_{N+1}=\overline{\mathbf{M}}\cdot\overline{\mathbf{v}}_N=(\overline{\mathbf{M}})^N\cdot\overline{\mathbf{v}}_1,
\label{eq:averagedequation}
\end{equation}
with $\overline{\mathbf{v}}_1=(1,1,1)^T$ as the boundary condition.
From Eq.~\eqref{eq:averagedequation}, $\overline{\mathbf{v}}_{N}$ can be obtained for arbitrary $N$. Substitution into the disorder averaged form of Eq.~\eqref{eq:XNrec} gives $\overline{X}_N$. This locates the phase transition: $\overline{X}_N$ is unity for all $N$ in the percolating phase, and decreases with increasing $N$ in the localised phase.

The critical disorder strength and localisation length exponent can be evaluated analytically by computing the eigenvalues of $\overline{\mathbf{M}}$.
We expect the largest eigenvalue, denoted by $\lambda_\mathrm{max}$, to be unity in the percolating phase and smaller than unity in the localised phase.
We indeed find~\cite{supp}
\begin{equation}
\lambda_\mathrm{max} = \begin{cases}1~~~~~~~~\mbox{for}~~W\le J/2\\
							\Lambda<1~~\mbox{for}~~ J/2 < W
				\end{cases}
\label{eq:lambdamax}
\end{equation}
with $\Lambda = \{J+4J_z+[(J-4 J_z) (8W -3 J-4 J_z)]^{\frac{1}{2}}\}/{4 W}$ for  ${J}/{2}\le W<{(J+4J_z})/{2}$.
This shows that the critical disorder strength is $W_c=J/2$.
The value of the exponent $\nu$ is determined by the asymptotic form
\begin{equation}
1 - \lambda_\mathrm{max}(W) \sim (W-W_c)^\nu ~~\mbox{for}~~W\gtrsim W_c.
\end{equation}
Expanding Eq.~\eqref{eq:lambdamax} around $W_c=J/2$ for $W>J/2$, we find
\begin{equation}
\lambda_\mathrm{max} = 1 - \frac{4}{J(J-4J_z)}(W-W_c)^2 + \mathcal{O}[(W-W_c)^3]\,.
\end{equation}
From this we identify the value $\nu=2$.
The analytical determination of $W_c$ and $\nu$ constitute two of our central results.

Understanding of the critical point is provided by a simple picture for the appearance of short, real-space segments in the chain where spins cannot fluctuate.
As an example, consider three consecutive sites, $\ell-1$, $\ell$ and $\ell+1$. Suppose that $\vert 2h_{\ell\pm1}\vert>J$ and $\vert 2h_\ell+4J_z\vert>J$. Then the spins at these sites in a Fock-space cluster with $\sigma^z_{\ell-1}= \sigma^z_\ell = \sigma^z_{\ell+1} = 1$  are frozen: the cluster has no active edges to nodes with other configurations of these spins, regardless of the orientation of the spins at sites $\ell\pm2$.
Such a disorder realisation requires fields on two end sites ($h_{\ell\pm1}$ in this case) exceeding a critical strength $W_c=J/2$. Similar arguments also apply if the end sites are adjacent, or if they have more than one site separating them. Frozen segments therefore appear with a density proportional to $(W-W_c)^2$ and their separation defines a correlation length, implying $\nu=2$.

\begin{figure}
\includegraphics[width=1\columnwidth]{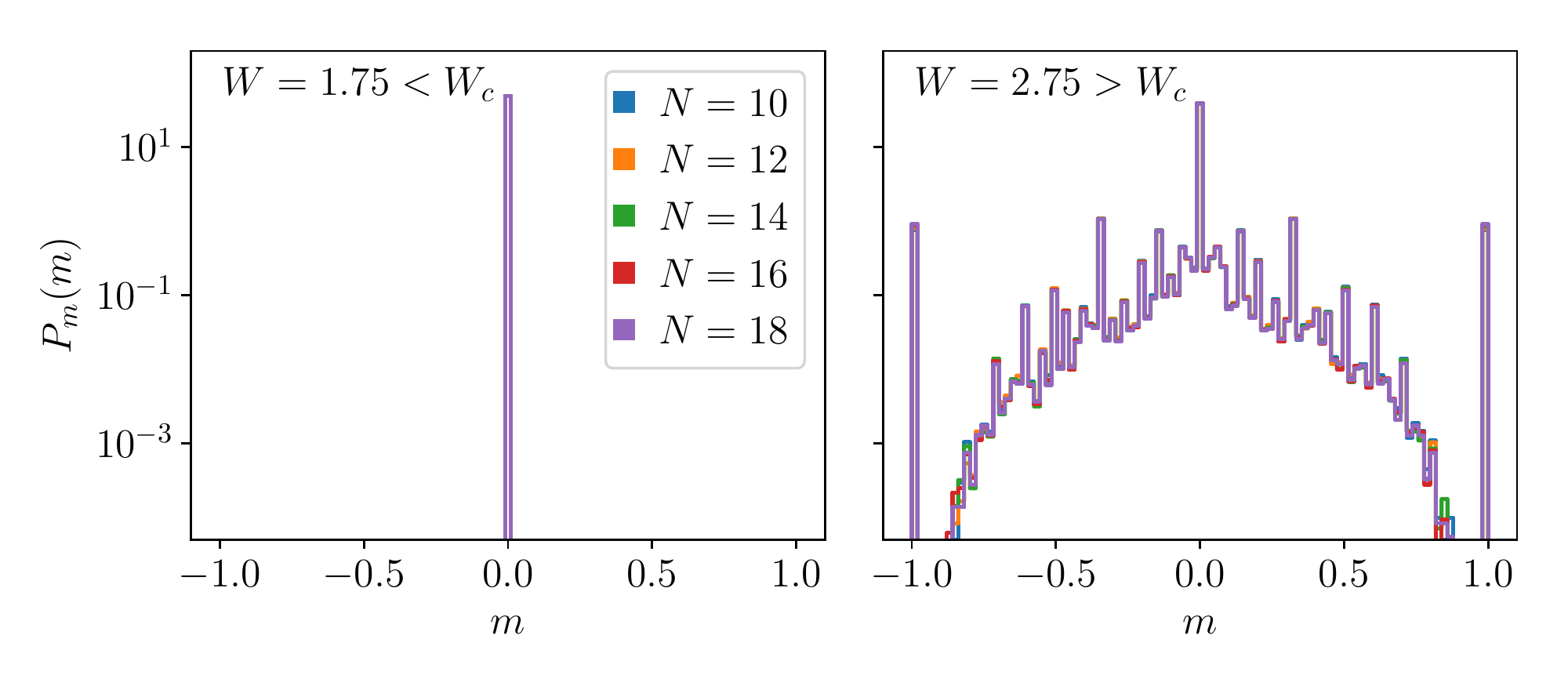}
\caption{Probability distribution of the cluster-averaged magnetisation in the percolating (left) and localised (right) phases, computed by complete enumeration of clusters.}
\label{fig:P(m)}
\end{figure}

We note that the value $\nu=2$ is consistent with the Harris-CCFS bound~\cite{harris1974effect,chayes1986finite}, which stipulates for a disorder-driven phase transition 
that $\nu\ge 2/d$ in spatial dimension $d$. A similar bound has been derived for many-body localisation transitions~\cite{chandran2015finite,khemani2017critical}. It is interesting that a scaling theory of entanglement at the many-body localisation transition, based on treating the many-body resonances classically in real space, also yields $\nu=2$ for the typical data~\cite{dumitrescu2017scaling}.

It is natural to ask how important a role the exchange interaction $J_z$ plays in the transition, particularly since $W_c$ is independent of $J_z$. In fact, the character of the localised phase is controlled by interactions. This is illustrated most directly by the fact that $\nu$ is discontinuous, taking the value $\nu=1$ at $J_z=0$~\cite{supp}. It is revealed in more detail by considering the magnetisation $m$ at a given real-space site, averaged over all nodes in a Fock-space cluster, and the probability distribution $P_m(m)$ of this quantity over disorder realisations. In the percolating phase $P_m(m)= \delta(m)$, while in the localised phase without interactions $P_m(m)$ has delta-function components at $m=0$ and $m=\pm 1$. As illustrated in Fig.~\ref{fig:P(m)}, interactions generate a broad background in addition to the delta-function components, which we study in detail elsewhere \cite{roy2018numerical}. Analogous differences in the distributions of eigenstate expectation values of local observables have also been observed across the many-body localisation transition~\cite{baldwin2016manybody,luitz2016long}.

In summary, we have constructed a classical percolation model in the Fock space of a disordered, interacting quantum spin chain. The classical model mimics aspects of the many-body localisation transition in the quantum system. Using a transfer matrix approach we have computed exactly for the classical problem the critical disorder strength and localisation length exponent. We have corroborated these results by enumerating clusters exactly on finite-sized systems.

A number of interesting directions remain open. In particular, the Fock-space percolation model can be interpreted as an 
example of a kinetically constrained model. Such models are known to show dynamical phase transitions and non-ergodic behaviour~\cite{sherrington2002glassy,garrahan2007dynamical,ritort2003glassy,garrahan2011kinetically}. They can be studied using Monte Carlo dynamics, which gives access to much larger system sizes than complete enumeration. This perspective also leads to the introduction of new, dynamical correlation functions, which we examine in separate work~\cite{roy2018numerical}.

\begin{acknowledgments}
We would like to thank S. Gopalakrishnan, A. Nahum, and S. A. Parameswaran for useful discussions. This work was in part supported by EPSRC Grant No. EP/N01930X/1.
\end{acknowledgments}

\bibliography{refs}

\clearpage

\onecolumngrid
\renewcommand{\arraystretch}{1.5}
\newcommand{\dfour}{{%
  \mathchoice{\includegraphics[height=2.5ex]{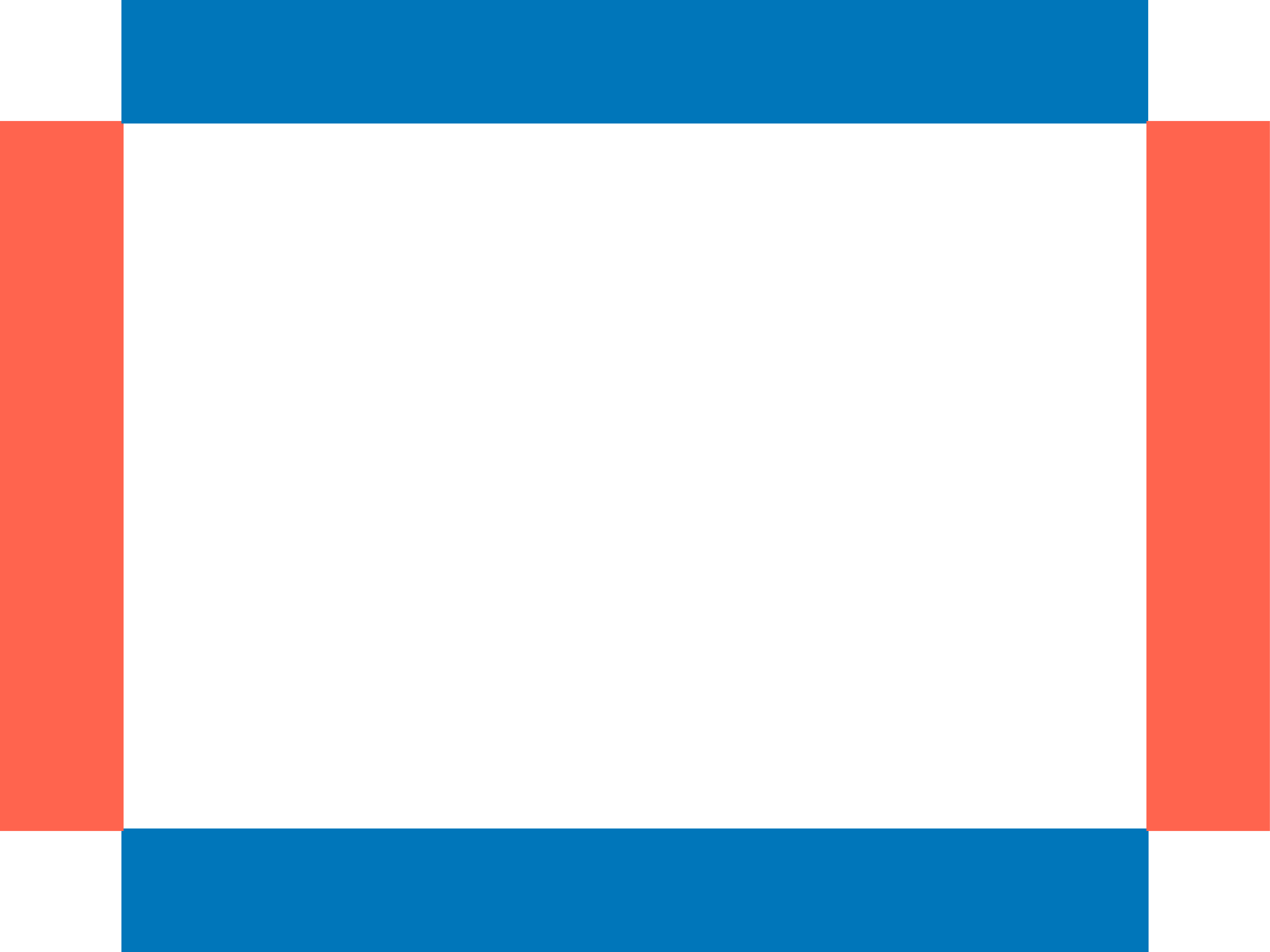}}
    {\includegraphics[height=2.5ex]{d4.pdf}}
    {\includegraphics[height=2.5ex]{d4.pdf}}
    {\includegraphics[height=2.5ex]{d4.pdf}}
}}

\newcommand{\dta}{\ensuremath{%
  \mathchoice{\includegraphics[height=2.5ex]{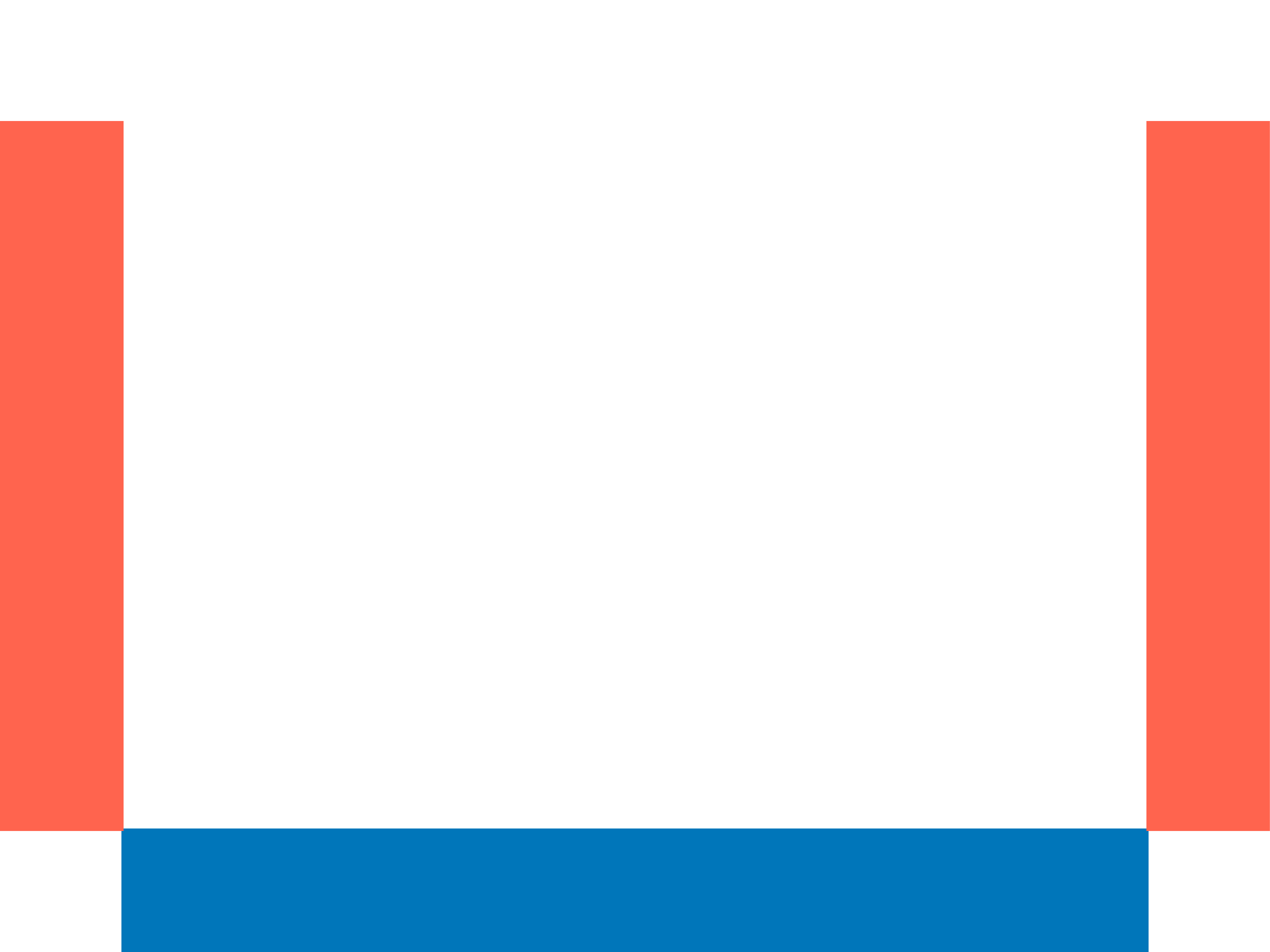}}
    {\includegraphics[height=2.5ex]{d3a.pdf}}
    {\includegraphics[height=2.5ex]{d3a.pdf}}
    {\includegraphics[height=2.5ex]{d3a.pdf}}
}}

\newcommand{\dtb}{\ensuremath{%
  \mathchoice{\includegraphics[height=2.5ex]{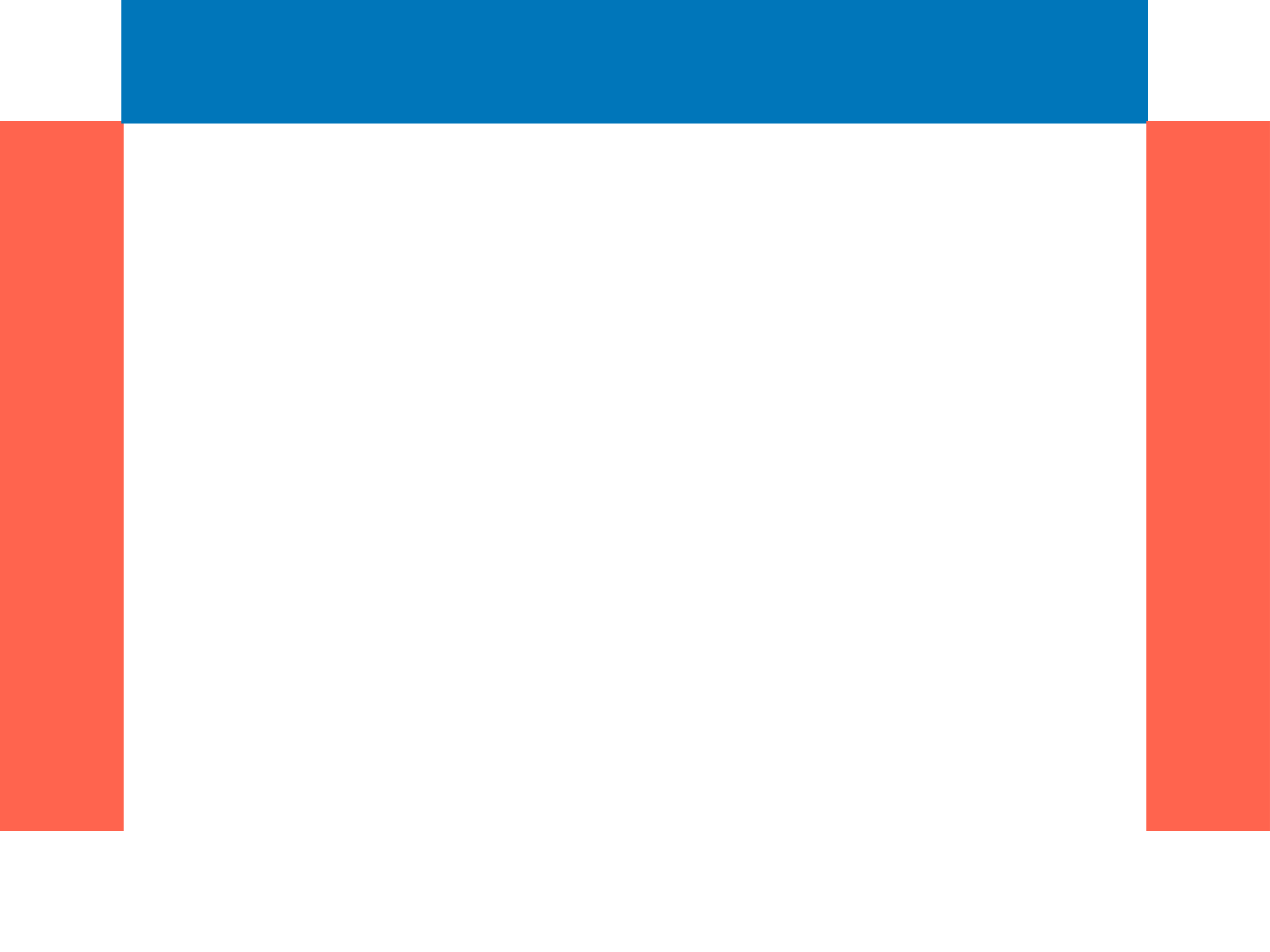}}
    {\includegraphics[height=2.5ex]{d3b.pdf}}
    {\includegraphics[height=2.5ex]{d3b.pdf}}
    {\includegraphics[height=2.5ex]{d3b.pdf}}
}}

\newcommand{\dtc}{\ensuremath{%
  \mathchoice{\includegraphics[height=2.5ex]{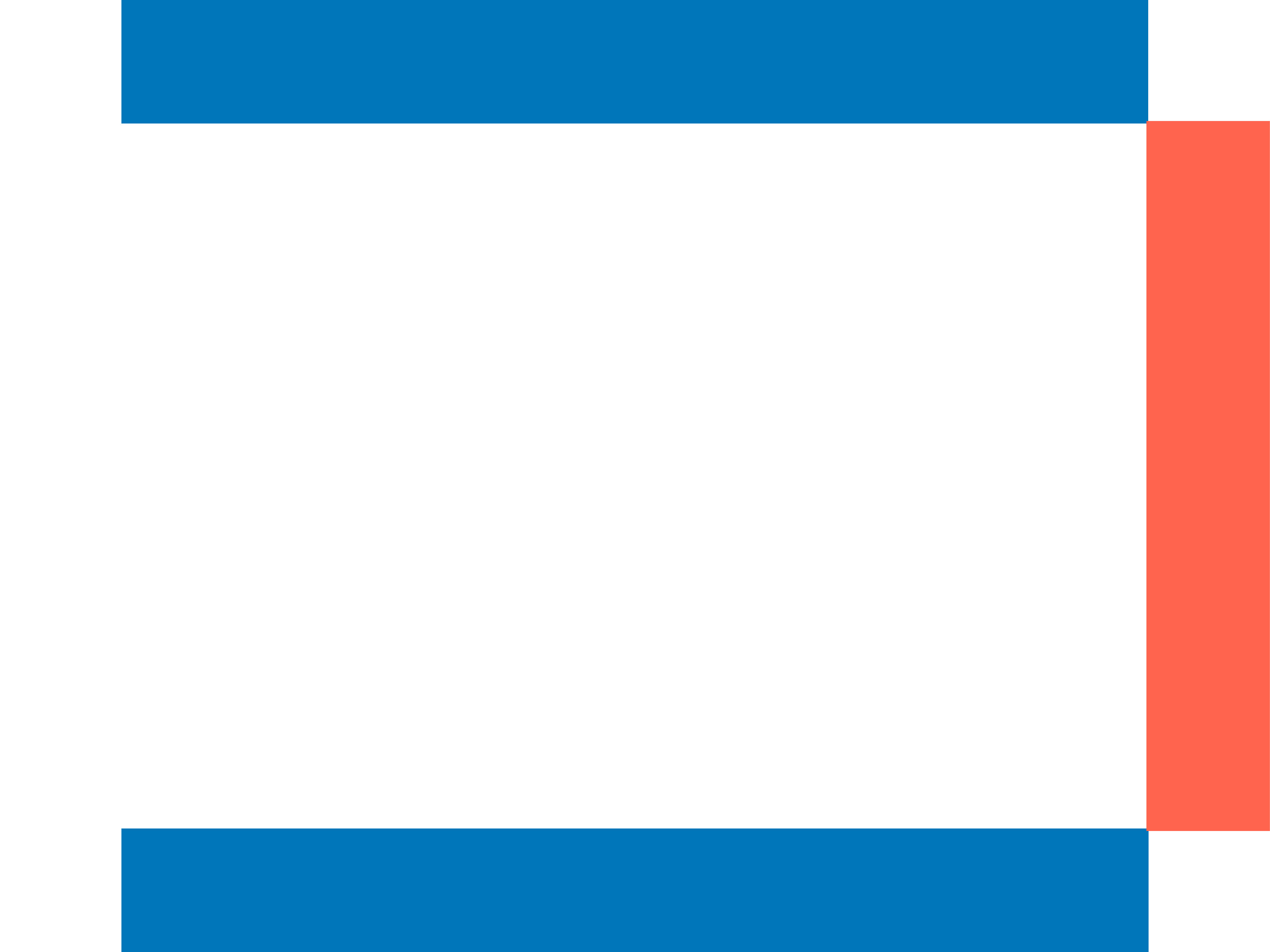}}
    {\includegraphics[height=2.5ex]{d3c.pdf}}
    {\includegraphics[height=2.5ex]{d3c.pdf}}
    {\includegraphics[height=2.5ex]{d3c.pdf}}
}}

\newcommand{\dtd}{\ensuremath{%
  \mathchoice{\includegraphics[height=2.5ex]{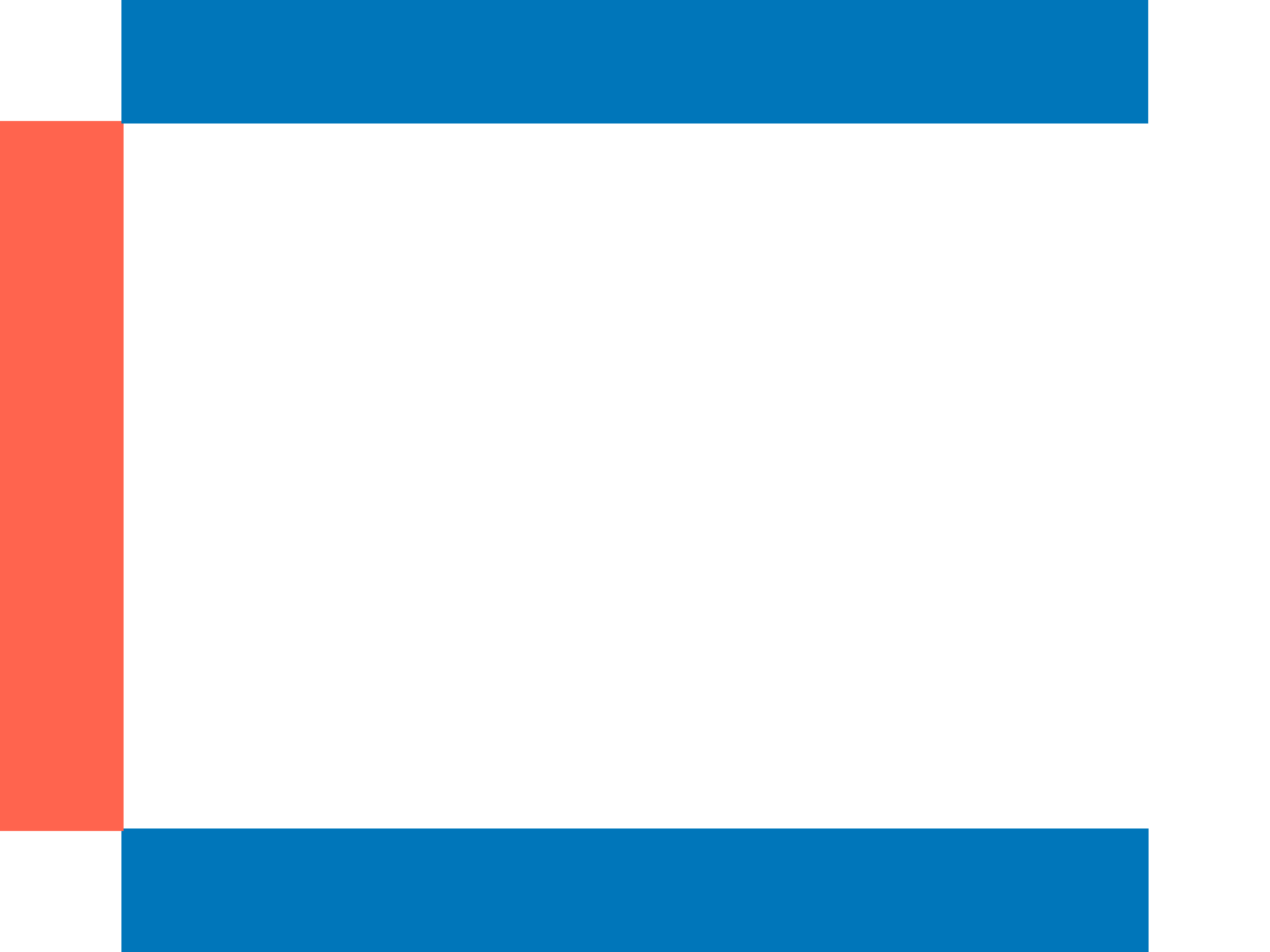}}
    {\includegraphics[height=2.5ex]{d3d.pdf}}
    {\includegraphics[height=2.5ex]{d3d.pdf}}
    {\includegraphics[height=2.5ex]{d3d.pdf}}
}}

\setcounter{equation}{0}
\renewcommand{\theequation}{S\arabic{equation}}
\setcounter{figure}{0}
\renewcommand{\thefigure}{S\arabic{figure}}

\begin{center}
{
\textbf{Supplementary material for ``Exact solution of a percolation analogue\\ for the many-body localisation transition''}\\
\medskip

Sthitadhi Roy\textsuperscript{1,2}, David E.~Logan\textsuperscript{1}, and J.~T.~Chalker\textsuperscript{2}\\
\smallskip

\small{\textsuperscript{1}\textit{Physical and Theoretical Chemistry, Oxford University,\\ South Parks Road, Oxford OX1 3QZ, United Kingdom}\\
\textsuperscript{2}\textit{Rudolf Peierls Centre for Theoretical Physics, Clarendon Laboratory,\\ Oxford University, Parks Road, Oxford OX1 3PU, United Kingdom}}
\medskip

}
\end{center}

This supplementary material presents details of our transfer matrix treatment of the Fock-space classical percolation problem that we have constructed from the many-body Hamiltonian for the disordered quantum Ising chain.

Our initial objective is to derive a recursion relation for  the indicator function $X_N(\{h_\ell\})$, which is defined for an open system of size $N$ and is a function of disorder configuration $\{h_\ell\}$. $X_N(\{h_\ell\})$ takes the value 1 if the all nodes of the Fock-space graph belong to the same cluster, and is zero otherwise. To express $X_{N+1}(\{h_\ell\})$ in terms of $X_N(\{h_\ell\})$ we refer to Fig.~\ref{fig:4Dstatespace}: in order that $X_{N+1}(\{h_\ell\})=1$, it is necessary that at least three of the four edges shown in this figure are active. This condition can be represented diagramatically in the form
\begin{equation}
X_{N+1} = \dfour~~+~~\dta~~+~~\dtb~~+~~\dtc~~+~~\dtd
\label{eq:sXNdiag}
\end{equation}
where the edges of the rectangles correspond to those in Fig.~\ref{fig:4Dstatespace}.
These diagrams denote the algebraic expressions
\begin{align}
\dfour~~&=X_N^+X_N^-x_{N+1}^+x_{N+1}^-, \nonumber\\
\dta~~&=X_N^+X_N^-x_{N+1}^+(1-x_{N+1}^-), \nonumber\\
\dtb~~&=X_N^+X_N^-(1-x_{N+1}^+)x_{N+1}^-,\nonumber\\
\dtc~~&=(1-X_N^+)X_N^-x_{N+1}^+x_{N+1}^-, \nonumber\\
\dtd~~&=X_N^+(1-X_N^-)x_{N+1}^+x_{N+1}^-,
\label{eq:diagex}
\end{align}
where, as in the main text, $X_N^\pm = X^\pd_N(h_1^\pd,\cdots,h_N^\pd\pm J_z^\pd)$ and $x_N^\pm = x(2h_N^\pd\pm 2J_z^\pd)$. 
Substituting the expressions given in Eq.~\eqref{eq:diagex} into Eq.~\eqref{eq:sXNdiag} yields Eq.~\eqref{eq:XNrec} of the main text, which is
\begin{align}
X^\pd_{N+1} = X_{N}^+X_{N}^-(x_{N+1}^+ + x_{N+1}^- -3x_{N+1}^+x_{N+1}^-)+(X_{N}^+ + X_{N}^-)x_{N+1}^+x_{N+1}^-.
\label{eq:sXNrec}
\end{align}
As $h_{N+1}$, the field at site $N+1$, enters Eq.~\eqref{eq:sXNrec} only via $x^\pm_{N+1}$, expressions for $X_{N+1}^\pm$ can be obtained simply by substituting $h_{N+1}\pm J_z$ in place of $h_{N+1}$ in the arguments of $x_{N+1}^\pm$ in Eq.~\eqref{eq:sXNrec}.
Some additional notation is useful for brevity. We define
\begin{itemize}
	\item[] $z_N^\pm(h_N^\pd) = x(2h_N^\pd\pm 4J_z^\pd) = x_N^\pm(h_N^\pd\pm J_z^\pd)$,
	\item[] $z_N^\pd(h_N^\pd) = x(2h_N^\pd)=x_N^\pm(h_N^\pd\mp J_z^\pd)$,
	\item[] $Y_N^\pd(\{h_\ell^\pd\}) = X_N^+X_N^-$.
\end{itemize}
Using this notation, the recursion relations for $X_{N+1}^\pm$ are
\begin{align}
X_{N+1}^\pm = Y_N^\pd(z_{N+1}^\pd+z_{N+1}^\pm-3z_{N+1}^\pd z_{N+1}^\pm)+(X_N^+ + X_N^-)z_{N+1}^\pd z_{N+1}^\pm.
\label{eq:sXNpmrec}
\end{align}
From Eq.~\eqref{eq:sXNpmrec} one can express $Y_{N+1}$ by multiplying $X_{N+1}^+$ and $X_{N+1}^-$. 
The functions $X_N$,  $X_N^\pm$, $Y_N$, $x$, $x^\pm$, $z$ and $z^\pm$ are idempotent, and in consequence $Y_N^\pd X_N^\pm = Y_N^\pd$. These properties lead to significant simplifications, making it possible to derive a closed set of linear recursion relations. 
Using these simplifications, we find
\begin{align}
Y_{N+1}^\pd = Y_N^\pd[z_{N+1}^\pd+z_{N+1}^+z_{N+1}^- - 3z_{N+1}^\pd z_{N+1}^+z_{N+1}^-] +(X_N^+ + X_N^-)z_{N+1}^\pd z_{N+1}^+z_{N+1}^-.
\label{eq:sYNrec}
\end{align}

Eqs.~\eqref{eq:sXNpmrec} and \eqref{eq:sYNrec} constitute a system of coupled linear recursion relations for $Y_N$, $X_N^+$, and $X_N^-$, of the form 
\begin{equation}
\begin{pmatrix}
Y^\pd_{N+1}\\X_{N+1}^+\\X_{N+1}^-
\end{pmatrix}=
\mathbf{M}^\pd_{N+1}\begin{pmatrix}
Y_{N}^\pd\\X_{N}^+\\X_{N}^-
\end{pmatrix}=
\begin{pmatrix}
a_{N+1}^\pd&&b_{N+1}^\pd&&b_{N+1}^\pd\\
c^+_{N+1}&&d^+_{N+1}&&d^+_{N+1}\\
c^-_{N+1}&&d^-_{N+1}&&d^-_{N+1}\\
\end{pmatrix}
\begin{pmatrix}
Y^\pd_N\\X_N^+\\X_N^-
\end{pmatrix},
\label{eq:mat}
\end{equation}
where 
\begin{eqnarray}
a_{N+1}^\pd &=&z_{N+1}^\pd+z_{N+1}^+z_{N+1}^- - 3z_{N+1}^\pd z_{N+1}^+z_{N+1}^-,\\
b_{N+1}^\pd &=&z_{N+1}^\pd z_{N+1}^+z_{N+1}^-,\\
c_{N+1}^\pm&=&z_{N+1}^\pd+z_{N+1}^\pm-3z_{N+1}z_{N+1}^\pm,\\
d_{N+1}^\pm&=&z_{N+1}^\pd z_{N+1}^\pm.
\end{eqnarray}
Since the matrix $\mathbf{M}_{N+1}$ depends only on the field $h_{N+1}$, the system of equations \eqref{eq:mat} can be converted to a system of recursion relations for the disorder averaged quantities.
We denote the disorder average by
\begin{equation}
\overline{(\cdots)}=\left(\prod_{\ell=1}^{N+1}\int dh_\ell P(h_\ell)\right)(\cdots),~~\mathrm{with}~~P(h) = \frac{1}{2W}\Theta(W-\vert h\vert),
\label{eq:hdist}
\end{equation}
where  $P(h)$ represents a uniform distribution for the disorder fields.

The disorder averaged system of recursion relation takes the form
\begin{equation}
\begin{pmatrix}
\overline{Y_{N+1}}\\\overline{X_{N+1}^+}\\\overline{X_{N+1}^-}
\end{pmatrix}=
\overline{\mathbf{M}}\begin{pmatrix}
\overline{Y_N}\\\overline{X_N^+}\\\overline{X_N^-}
\end{pmatrix}=
\begin{pmatrix}
\overline{a}&&\overline{b}&&\overline{b}\\
\overline{c^+}&&\overline{d^+}&&\overline{d^+}\\
\overline{c^-}&&\overline{d^-}&&\overline{d^-}
\end{pmatrix}
\begin{pmatrix}
\overline{Y_{N-1}}\\\overline{X_{N-1}^+}\\\overline{X_{N-1}^-}
\end{pmatrix},
\label{eq:matavg}
\end{equation}
with the boundary condition
\begin{equation}
\left(\overline{Y_{1}^\pd},\overline{X_{1}^+},\overline{X_{1}^-}\right) = (1,1,1).
\end{equation}
The averaged matrix elements in Eq.~\eqref{eq:matavg} can be calculated explicitly using Eq.~\eqref{eq:hdist}.
We find that there are four regimes of $W$ in which the averages are piecewise analytic, obtaining
\begin{center}
\begin{tabular}{|c||c|c|c|c|}
&$0<W\le \frac{J-4J_z}{2}$&$\frac{J-4J_z}{2}\le W\le \frac{J}{2}$&$\frac{J}{2}\le W<\frac{J+4J_z}{2}$&$\frac{J+4J_z}{2}\le W$\\
\hline
$\overline{a}$&$-1$ & $1-\frac{J-4J_z}{W} $& $\frac{-J+8J_z}{2W}$& $\frac{-J+8J_z}{2W}$\\
$\overline{b}$& 1& $\frac{J-4J_z}{2W}$& $\frac{J-4J_z}{2W}$& $\frac{J-4J_z}{2W}$\\
$\overline{c^\pm}$&$-1$ & $\frac{-J+4J_z}{2W}$&$\frac{-3J+8J_z+2W}{4W}$ & $\frac{-J+6J_z}{2W}$\\
$\overline{d^\pm}$& 1& $\frac{J-4J_z+2W}{4W}$&$\frac{J-2J_z}{2W}$ & $\frac{J-2J_z}{2W}$
\end{tabular}
\end{center}

For $W<(J-4J_z)/2$, substituting $\overline{a}=-1=\overline{c^\pm}$ and $\overline{b}=1=\overline{d^\pm}$ into Eq.~\eqref{eq:matavg}, one finds that $(\overline{Y_{N}},\overline{X_{N}^+},\overline{X_{N}^-}) = (1,1,1)$ is an eigenvector of $\overline{\mathbf{M}}$ with eigenvalue 1. In addition, in this regime $\langle x_N^\pm\rangle=1$ and hence $\overline{X_N}=1$ is the solution of Eq.~\eqref{eq:sXNrec}. This is the signature of a percolating phase. 
In contrast, for $W>(J+4J_z)/2$ we find $\lim_{N \to \infty}(\overline{\mathbf{M}})^N =0$ so that
$(\overline{Y_{N}},\overline{X_{N}^+},\overline{X_{N}^-})$ and $\overline{X_N}$ approach zero at large $N$, indicating a localised phase.
This shows that for $W<(J-4J_z)/2$ and $W>(J+4J_z)/2$, the system is definitely percolating and localised respectively.

To study the transition itself, the regime $(J-4J_z)/2<W<(J+4J_z)/2$ needs to be investigated.
In particular,
we require the largest eigenvalue of $\overline{\mathbf{M}}$. Its form (also given in the main text) is
\begin{equation}
\lambda_\mathrm{max} = \begin{cases}1;~~W<J/2\\
							\frac{\sqrt{(J-4 J_z) (-3 J-4 J_z+8 W)}+J+4 J_z}{4 W};~~ J/2\le W<(J+4J_z)/2\,.
				\end{cases}
\label{eq:slambdamax}
\end{equation}
This shows that the critical disorder strength is $W_c=J/2$. 
Expanding around $W_c$, we find
\begin{equation}
\lambda_\mathrm{max} = 1 - \frac{4}{J(J-4J_z)}(W-W_c)^2 + \mathcal{O}[(W-W_c)^3]\,.
\end{equation}
We identify from this the exponent value $\nu=2$. Note that, in the limit of $J_z=0$, the range of validity of the second line of Eq.~\eqref{eq:slambdamax} vanishes. 
In this case, it can easily be shown (using the expressions in the last column of the above table) that 
$\lambda_\mathrm{max} = J/2W$ for $W>J/2$ and $1$ for $W\le J/2$. Expanding in $W$ around $W_c=J/2$, 
$\lambda_\mathrm{max} = 1 - (2/J)(W-W_c) + \mathcal{O}[(W-W_c)^2]$, thus identifying $\nu=1$ in this limit.

\end{document}